\documentstyle[11pt,epsf]{article}

\begin{document}

\begin{center}

{\Large $\int e^{-x^2} dx$ and the Kink Soliton}

{\em Bruce Bassett}\footnote{email: bruce@stardust.sissa.it}

International School for Advanced Studies, \\
Via Beirut 2-4, 34014 Trieste, Italy\\

\date{\today}
\end{center}


\begin{abstract}
We provide analytical functions approximating $\int e^{-x^2} dx$, the 
basis of which is the kink soliton and which are both accurate (error $< 
0.2 \%$) and  simple. We demonstrate our results with some applications,  
particularly to the generation of Gaussian random fields. 
\end{abstract}

\section{Introduction}

There is an inherent asymmetry between integration and differentiation 
which makes integration somewhat of an art form, and which is perhaps 
best exemplified by the  lack of an 
elementary indefinite integral of the celebrated Gaussian:
\begin{equation}
\int \exp \left(\frac{-(x-\beta)^2}{\sigma^2}\right) dx
\label{eq:gauss}
\end{equation}
The fact that such an integral does not in fact exist follows from the work of 
Laplace \cite{rosen72,rosen68}. However, the Gaussian integral is 
fundamental, finding applications in statistics, error theory  
and many branches of physics. In fact, anywhere  one has  Gaussian 
distributions, cummulatives of these distributions will involve the above 
integral. Only special case  definite integrals of $e^{-x^2}$ are known, 
the most famous being: 
\begin{equation}
\int_0^{\infty} e^{-x^2/\sigma^2}dx = \frac{\sqrt{\pi} \sigma}{2}
\label{eq:infint}
\end{equation} 
In addition there is the series expansion \cite{GR80}:
\begin{equation}
\int_0^x e^{-u^2} du = \sum_{k = 1}^{\infty} \frac{(-1)^{k-1}}
{(k-1)!(2k - 1)} x^{2k - 1}
\label{eq:series1}
\end{equation}
Now in practise one can evaluate the integral accurately by numerical 
methods or tables, but in many  cases it would be preferable to have an 
analytical solution, even if it were not exact, as long as the maximum 
error were very small and the approximation were simple \footnote{Several 
rational function approximations exist but they are rather complicated 
\cite{AS65}.}. 

It turns out that there exists a function well known in the analysis of nonlinear 
partial differential equations whose derivative is very close to Gaussian - the kink 
soliton: 
\begin{equation}
\phi(x) \equiv  A \tanh(bx - c\beta)
\label{eq:kink}
\end{equation} 
with derivative:
\begin{equation}
\chi(x) \equiv Ab \left(1 - \tanh^2(bx - c\beta)\right)
\label{eq:family}
\end{equation} 
where $A,b,c,$ and $\beta$  are all real constants. 
The graphs of $e^{-x^2}$ and $\chi(x)$ are shown in figure (1). 
\footnote{One can consider a one-parameter family of approximations to the 
Gaussian given by replacing $x \rightarrow x^{\epsilon}$ in eq. 
(\ref{eq:family}) which give better fits when $\epsilon \not = 1$, but which do 
not have indefinite integrals as far as is known to the author.} 
The kink soliton is the positive, time-independent, topological 
solution to the non-linear $1+1$ dimensional partial differential 
equation: 
\begin{equation}
\phi_{tt} - \phi_{xx} = 2 b^2(\phi - \frac{1}{A^2}\phi^3) 
\end{equation}
where  a subscript denotes partial derivative 
with respect to that variable. The solution to this equation is
topological because the boundary conditions at $x = \pm \infty$ 
are different. 
Leaving the physical origin of 
$\phi$ behind, it is interesting to examine the series expansion of 
$\tanh(x)$:
\begin{equation}
\tanh(x) = \sum_{k=1}^{\infty} \frac{2^{2k} (2^{2k} - 1)}{(2k)!} B_{2k} 
x^{2k - 1}~,~~~~\mbox{valid for}~~x < \frac{\pi}{2}
\label{eq:series2}
\end{equation} 
which should be compared with eq. (\ref{eq:series1}) for $\int e^{-x^2} dx$.
Here $B_{k}$ are the Bernoulli numbers with 
generating function $t/(e^t -1)$. We see that although the coefficients 
differ in each case, the powers of $x$ in the expansions are identical.  
Further both $\chi(x)$ and $e^{-x^2}$ have the property that their 
derivatives can be re-expressed in terms of themselves and $\phi(x)$ or 
powers of $x$ respectively. These observations shed some light on the 
foundations of the approximation.  

\begin{figure}[hbp]
\epsffile[50 50 410 302]{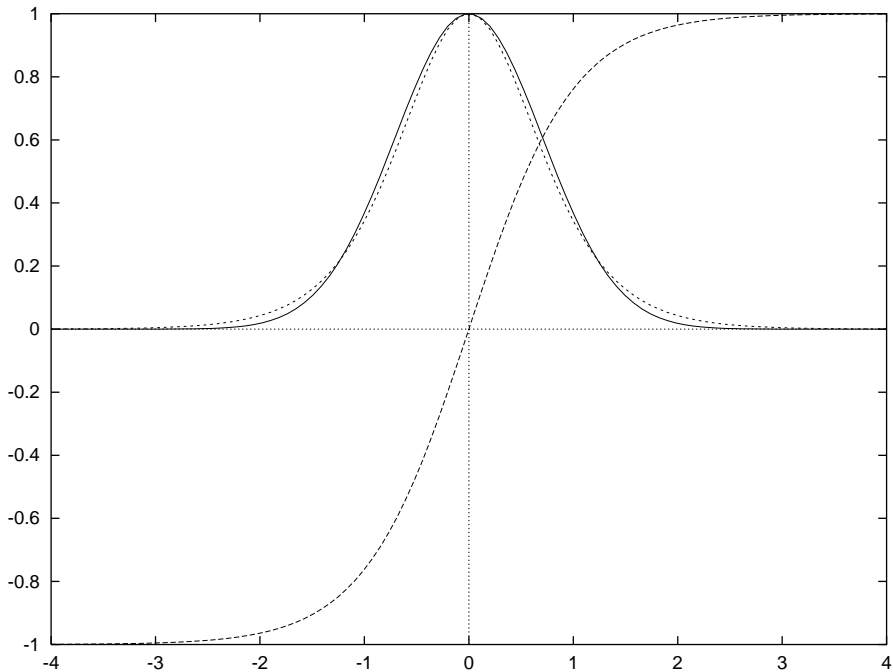}
\caption{Plot of $e^{-x^2}$ (solid line), $\chi(x)$ (dotted line) and
$\tanh(x)$ (dashed line) which is the kink soliton.}
\label{fig:exp}
\end{figure}

\section{Details of the approximation}
Turning to practical issues, we are 
left with choosing the constants, $A,b,c$ to optimise the approximation 
of eq. (\ref{eq:gauss}). 
We need three constraints  to fix the three parameters. First we require 
that the Gaussian and $\chi(x)$  have the same symmetry axis. This 
requires the argument of  $\tanh$ to vanish at $x_* = \beta$ which 
immediately implies from eq. (\ref{eq:family}) that $c = b$. 

At this stage we have a choice, dependent on whether we are interested in 
an approximate solution for  small or large $x$. For large $x$, a 
constraint is obviously that our new 
approximation, $\phi(x)$, must give {\em exactly} the same result as  eq. 
(\ref{eq:infint}) when differenced at 
infinity and the origin. This will ensure  convergence of our 
approximation. Since $\tanh(x) \rightarrow 1$ 
as $x \rightarrow \infty$, and $\tanh(0) = 0$, this implies from 
eq. (\ref{eq:kink}) that: \[ A = \frac{\sqrt{\pi}\sigma}{2}\]

Finally we can impose that $\chi(x) = e^{-(x-\beta)^2/\sigma^2}$ at 
some point, i.e, we  match the derivatives. We will choose $x = \beta$ 
as the simplest. 
This gives: \[ Ab = 1 \Longrightarrow b = \frac{2}{\sqrt{\pi}\sigma} \]
In fact the two are equal at another point as can be seen from 
figure (1).
Our analytical approximation, which is very accurate for large $x$, 
is therefore: 
\begin{equation}
\phi(x) = \frac{\sqrt{\pi}\sigma}{2} \tanh \left( \frac{2}{\sqrt{\pi}\sigma}(x - 
\beta)\right) ~\simeq  \int e^{-(x-\beta)^2/\sigma^2} dx
\label{eq:integral}
\end{equation}
where in this paper $\simeq$ is understood as meaning asymptotic  
convergence, as $x \rightarrow \infty$ and bounded error $\forall x$.
From figures (1,2) we see that the kink derivative underestimates the 
Gaussian 
at small $(x - \beta)^2$ and overestimates it at large $(x - \beta)^2$. 

\begin{figure}[!hbp]
\epsffile[50 50 410 302]{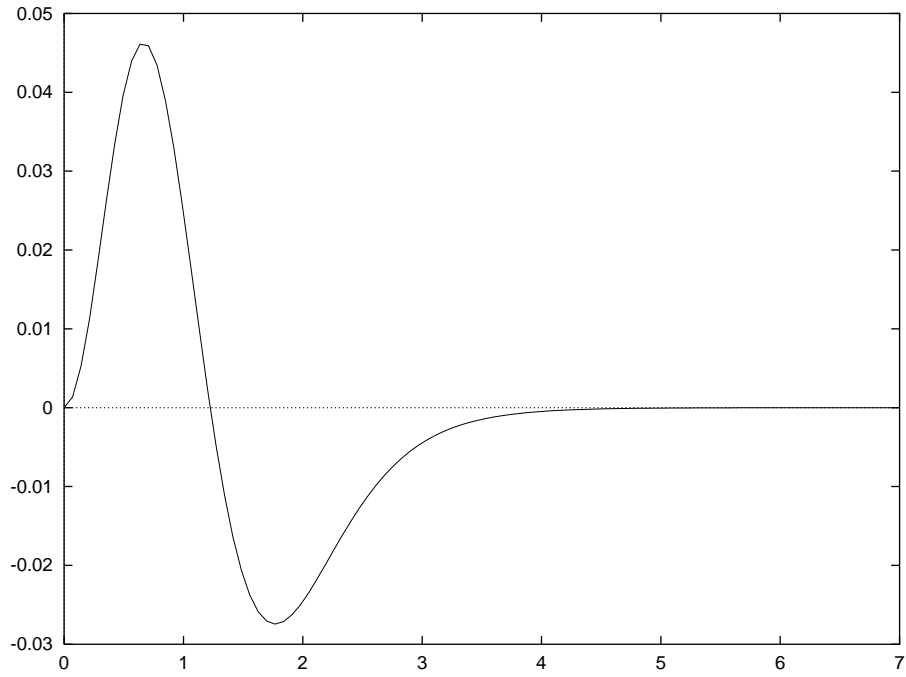}
\caption{Plot of the difference between $e^{-x^2}$ and $\chi(x)$.}
\label{fig:diff}
\end{figure}

Alternatively if one is interested in  $\int_0^u
e^{-x^2/\sigma^2}dx$ where $u \le 4\sigma$ say, then this will not be
good enough, since the error in our approximation is strongly confined to 
small $x$.
Instead we can impose that $\phi(x)$  must give the exact 
result, not at infinity, but at the end of the interval, i.e. at $u$. 
Thus we impose: 
\begin{equation}
A \tanh(b(u - \beta)) = \int_0^u e^{-(x-\beta)^2/\sigma^2} dx
\label{eq:smallx}
\end{equation}
In addition we need to match the derivatives $\chi(x_*) = 
e^{-(x_*-\beta)^2/\sigma^2)}$ at some point $x_*$ as before,  and then solve 
the equations for $A,b$. It is an open question  which matching 
point yields 
the best results. For illustrative purposes we choose $x = \beta$ and  
again find $A = 
1/b$, so that substituting in eq.(\ref{eq:smallx}) gives us a nonlinear 
root-finding problem for $A$. The right-hand side can be found for 
example, from tables of the error function, $\mbox{erf}(x)$. This yields an 
approximation which is exact at $x = u$ and hence a much better 
approximation for small $x$,  but which is invalid for $x \gg u$.  
The extension to cases with variable lower limit of integration is obvious 
and will not be considered.

\begin{figure}
\epsffile[50 50 410 302]{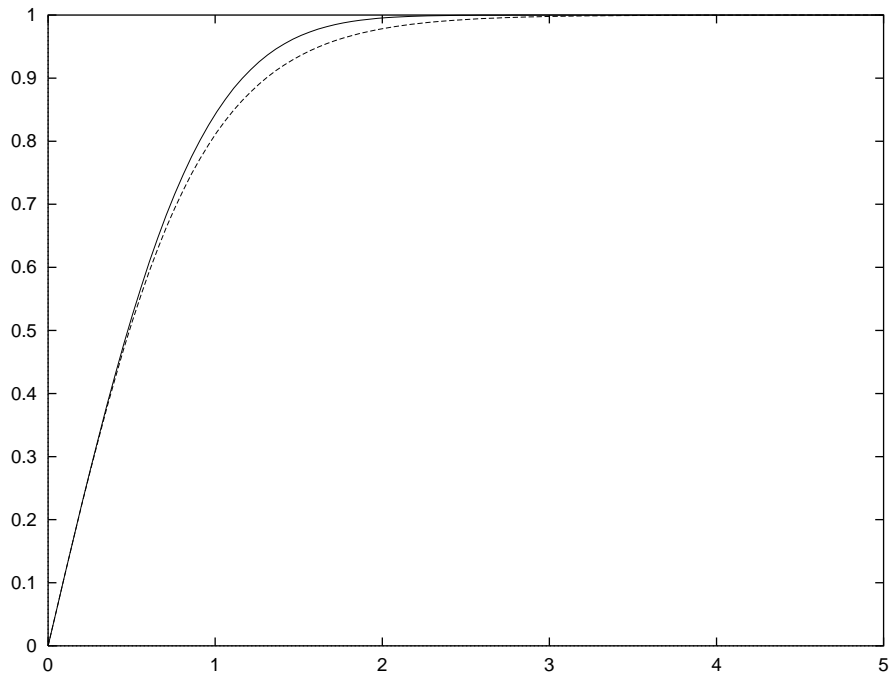}
\caption{Plot of the error function, 
and the soliton approximant, $\phi(x)$. The maximum difference 
occurs at $x = 1.12$ and is $3.91\%$. The error drops below $1\%$ 
for $x \ge 2.3$ and converges exponentially to zero.}
\label{fig:integral}
\end{figure}

One might be tempted to generalise eq. (\ref{eq:kink}) to a one-parameter family of 
approximations to the error function: 
\begin{equation}
\Delta_p(x) = A \tanh^p(bx)
\label{eq:tanp}
\end{equation}
which have derivative:
\begin{equation}
\Delta_p'(x) = Ab p \tanh^{p-1}(bx) \mbox{sech}^2(bx) 
\label{eq:derivdel}
\end{equation}
However, since for $p \not = 1$, $\Delta_p'(0) = 0$, they are not really 
suitable as 
approximations to a Gaussian. Rather they are skewed distributions with 
maxima at $x > 0$. It turns out however, that they will be useful later.

For testing our approximation we will use the $\phi(x)$ valid for large 
$x$, denoted $\phi(x)_L$, given by eq. (\ref{eq:integral}). The crucial 
question is of 
course, how good is this approximation ? It turns out that it is very good 
in most cases, as can be seen from figures (3) and (4). The 
maximum error from using $\phi(x)_L$ is
$3.91 \%$ at $x = 1.12$. However
as discussed earlier, if one 
is interested in the result for small $x$, and $x_1$ is small, then this 
is not the best approximation to use. In practise, the 
error drops off very 
quickly due to the exponential nature of $\tanh(x)$. For example, 
the  error in estimating $\mbox{erf}(x)$ drops below $1\%$ for $x \ge 
2.3$ and at $x = 
5$ the error is  $2.51 \times 10^{-5} $. The error as a function of 
x is plotted in figure (4). 

\section{Improving the approximation}

The shape of figure (4) is, in fact, rather 
startling because it is a very simple  shape. From the graph it  has a 
single local maximum and hence two points where the concavity changes.  
Hence although  it cannot be written down explicitely in terms of 
elementary functions  \cite{rosen68}, it 
can be approximated very closely. Several fitting shapes were tried, 
such as the 
log-normal and Poisson distributions, but the best was found to be a  
generalised Maxwell-distribution: 
\begin{equation}
E(x) = \alpha_1 x^n \exp(-\frac{x^2}{\alpha_2}) 
\label{eq:differr}
\end{equation}
For the case used in the figures, that of $\mbox{erf}(x)$, the best 
parameters for reducing the maximum error (i.e. minimising w.r.t. the 
sup-norm $||\cdot||_{\infty}$) were (see figure (5)): 
\begin{equation}
\alpha_1 = 0.062~,~~~n = 2.27~,~~~\alpha_2 = 1.43
\label{eq:param}
\end{equation}
which reduced the {\em maximum} error to  $0.15 \%$. 
It is also likely  that our choice of function and parameters for $E(x)$ is not 
optimal, since  formal optimisation was not used, but was based rather on a 
numerical investigation of the parameter space $\{\alpha_1,\alpha_2,n\}$.

Further, since the required  $E(x)$ is a skewed Gaussian with maximum at non-zero $x$ 
we can profitably  employ the functions given by eq. (\ref{eq:derivdel}), 
originally introduced to model the Gaussian, as fits for the error. In this 
case our approximation becomes: 
\begin{equation}
\int_0^x e^{u^2}du = \frac{\sqrt{\pi}}{2} \left[ \tanh(\frac{2}{\sqrt{\pi}}x) + 
(\alpha_3 \tanh^p(x))' \right]
\label{eq:tot}
\end{equation}
where $'$ denotes derivative w.r.t. $x$. For $\alpha_3 = 0.23$ and $p = 9.7$ 
the error is at most $9 \times 10^{-3}$. By suitable generalisation of the 
second term it is possible to increase the accuracy to the level of the 
generalised Maxwell distribution, but for simplicity and because of its 
suggestiveness, we leave it in the above form.

\begin{figure}[!hbp]
\epsffile[50 50 410 302]{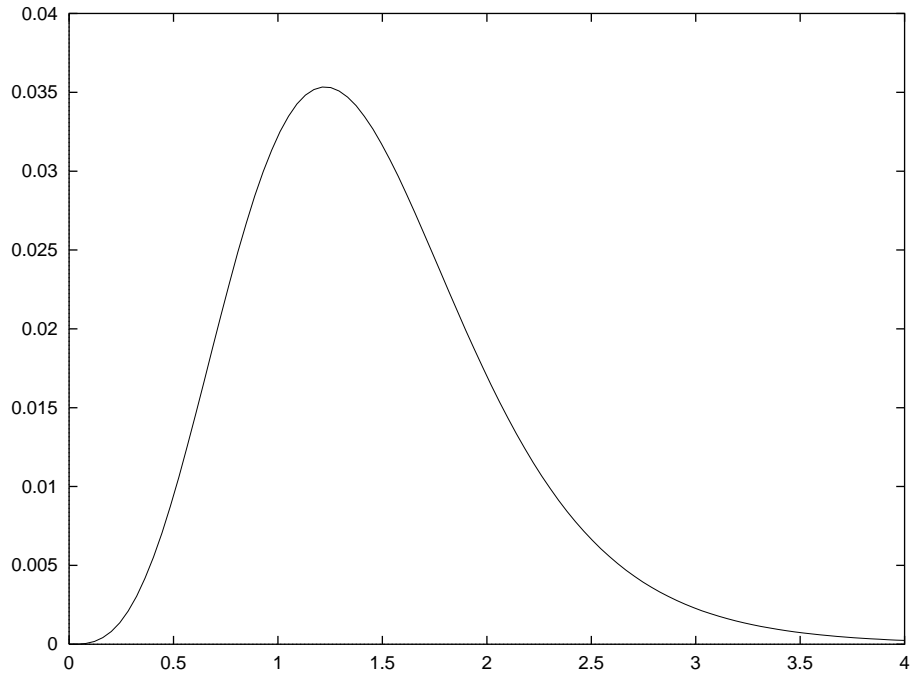}
\caption{The difference of $\mbox{erf}(x)$ and $\phi(x)_L$. This is 
closely approximated
by  log-normal distributions or generalised Maxwellians  of the form 
$\alpha_1 x^n e^{-x^2/\alpha_2}$. }
\label{fig:diff2}
\end{figure}

\begin{figure}
\epsffile[50 50 410 302]{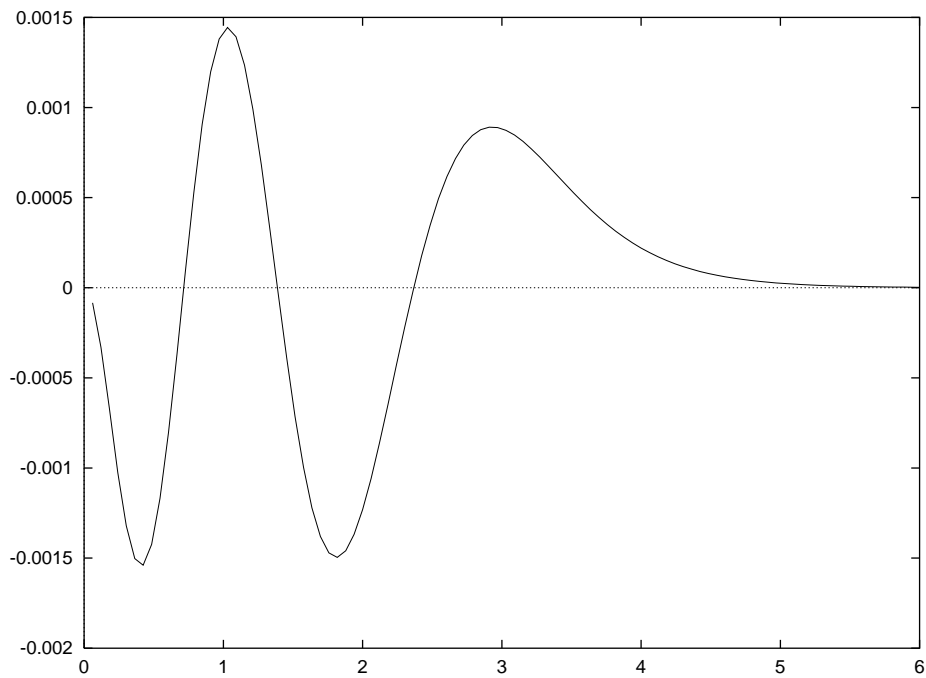}
\caption{The  final error after modeling figure (4) by the generalised Maxwell 
distribution of eq.(\ref{eq:differr}). The maximum error is about $0.15\%$.}
\label{fig:reserror}
\end{figure}

In the case of the error function we have explicitely that
($\beta = 0$): 
\begin{equation}
\mbox{erf}(x) \simeq  \tanh(\frac{2}{\sqrt{\pi}}x) + E(x)
\label{eq:erf}
\end{equation} 
where $\mbox{erf}(x) \equiv \Phi(x)  = 2/\sqrt{\pi} \int_0^x e^{-u^2} du$ is 
the error function. Similarly the complementary error 
function is given by: $\mbox{erfc}(x) = 1 - \tanh(\frac{2}{\sqrt{\pi}}x) 
- E(x)$.

\section{Moments of the soliton}

A fundamental feature of a Gaussian distributed random variable is that 
all moments above the second, such as the skewness, are zero. From this it 
follows that the sum of error distributed random variables is itself 
error distributed. A natural question to ask is how well the soliton 
approximation preserves this feature.

To make this more precise: given the distribution $P(x)$, we may define the 
partition function $Z(J)$ 
\footnote{We use the notation $Z(J)$ because of its ubiquitous use 
in statistical physics. In the case where $x$ is a function, $Z(J)$ 
becomes a 
path-integral and derivative becomes functional derivative in eq. 
(\ref{eq:moment}).} via: \begin{equation} 
Z(J) = \int P(x) e^{J x} dx
\label{eq:part}
\end{equation}
From the ``free energy" $F(J) = \ln Z(J)$ we may now define the 
n-th moment, $M_n$, of $P(x)$ as:
\begin{equation}
M_n \equiv \frac{d^n}{d J^n} F(J)|_{J = 0}
\label{eq:moment}
\end{equation}

Thus in  the case of a Gaussian distribution  with zero mean, it is easy to 
show that the free energy is a quadratic function of $J$. Hence the only 
non-zero moment is the second, i.e. the variance, as claimed above. In the 
case of the soliton approximant we have:
\begin{equation}
Z(J) = \int [1 - \tanh^2(2x/\sqrt{\pi} \sigma)] e^{J x} dx
\end{equation}
which is unfortunately not known analytically, so we resort to numerical 
analysis. Using 
the Gaussian case as a testbed we approximated the free energy with an 
8-th degree polynomial:
\begin{equation}
F(J) \sim \Sigma_{n = 0}^8 \alpha_n J^n
\label{eq:poly}
\end{equation} 
For a Gaussian $\alpha_n = 0, ~~n \ge 3$. Using a least-squares method, the 
error, i.e. the largest $\alpha_n$ coefficient which is zero in the exact case 
but non-zero in the fit, was $\alpha_3 = 2.535 \times 10^{-8}$.
Each subsequent coefficient was roughly an order of magnitude smaller 
than the preceding one.

In the case of the soliton approximation, given by eq. (\ref{eq:family}), 
the error was  $2.309 \times 10^{-3}$  again for the cubic term, and 
again with roughly $\alpha_{n+1} \sim \alpha_n/10$. 

\vspace*{0.5cm}
\begin{center}
\begin{tabular}{|c|c|c|}
\hline
\hline
 $\alpha_3$ & $\alpha_4$ & $\alpha_5$ \\ 
\hline
\hline
 $-2.535 \times 10^{-8}$ &  
 $4.202 \times 10^{-9}$   
& $-4.002 \times 10^{-10}$ \\
 $-2.309 \times 10^{-3}$ &  
 $4.308 \times 10^{-4}$ &  
 $-4.389 \times 10^{-5}$ \\
\hline
\hline
$\alpha_6$ & $\alpha_7$ & $\alpha_8$ \\
\hline
\hline
 $2.182 \times 10^{-11}$ &
 $-6.322 \times 10^{-13}$ &
 $7.532 \times 10^{-15}$ \\
 $2.586 \times 10^{-6}$ &
 $-8.144 \times 10^{-8}$ &
 $1.071 \times 10^{-9}$ \\
\hline
\hline
\end{tabular}
\end{center}
\vspace*{.5cm}

Table (1) shows a comparison between the coefficients of the free-energy 
polynomials for the terms higher than cubic for the exact Gaussian and the 
soliton approximant $\chi(x)$. An interesting thing to note is that, 
although the accuracy is at the level one might expect, i.e. $\sim 
10^{-3}$, the pattern of 
the terms is identical; namely both the signs and the decrease in the 
coefficients have the same behaviour in both cases. This suggests that 
numerical errors will be ``coherent", i.e. the errors one has from 
numerical integration of the Gaussian will be of the same nature as those one 
obtains from the soliton approximation. This is perhaps obvious given the 
similarity of their power series (see eq.s 
(\ref{eq:series1},\ref{eq:series2})) but will not be true for other 
approximants in terms of e.g. rational functions \cite{AS65}.

We leave this discussion by noting that inclusion of $E(x)$, via e.g. 
eq. (\ref{eq:differr}),  in 
the  calculation of moments will reduce the above  errors considerably, 
presumably by a factor of at least $10^2$.

\section{Applications}

Let us now consider a  small sample  of   applications. 
A primary example is in the theory of statistics. If we have a  
uniformly distributed random variable $\chi$ and we desire a random 
variable $y$ with statistics given by a distribution $f$, first define   
the integral $F(x) = \int_0^x f(\chi) d\chi$. Then 
$y = F^{-1}(x)$ will have the same distribution 
as $f$, where $F^{-1}$ denotes the inverse of $F$, on 
the interval $[F^{-1}(0),F^{-1}(x)]$. 

In particular if, as is often the case, we want to generate a realisation of a 
Gaussian random distribution, $f =  \exp(-x^2/\sigma^2)$, then with our 
approximation, $F(x) = \phi(x)$ (we have dropped the error correction term 
$E(x)$ for simplicity) and the inverse $\phi^{-1}(x)$, gives us our random 
variable. In this case if $y = \phi(x)$, then:
\begin{equation}
\phi^{-1}(x) = \frac{\sqrt{\pi} \sigma}{2} 
\tanh^{-1}(\frac{2}{\sqrt{\pi}\sigma} x) 
\end{equation} 
which has the same form as $\phi(x)$ with the replacement 
$\tanh \rightarrow \tanh^{-1}$ so that both the integral and inverse are 
essentially trivial. This avoids the necessity of using traditional Monte 
Carlo methods to calculate Gaussian distributions.

A related problem occurs in the study of structure formation from 
gravitational collapse from Gaussian initial conditions, a standard
assumption. The Press-Schecter formalism \cite{PS74}, gives the 
cummulative mass function $f(>M)$, which is the number of objects (such 
as galaxies) with mass greater than $M$: 
\begin{equation}
f(\ge M) = 1 - \mbox{erfc} \left( \frac{\delta_c}{\sqrt{2} \sigma(M,z)} \right )
\label{eq:ps}
\end{equation}
where $\delta_c,z \in {\boldmath R}$ and $\sigma$ is the variance of the 
distribution. This can be estimated immediately  using eq. 
(\ref{eq:erf}).

One place where error functions are ubiquitous is in diffusion theory, since 
the decaying Gaussian is a solution to the standard 
diffusion equation. In the case where there is an extended distribution of 
diffusing material, situated at  $x < 0$ for example, the solution is instead 
given by: 
\begin{equation}
C(x,t) = \frac{C_0}{2} \mbox{erfc} \left(\frac{x}{2\sqrt{ D t }} \right) 
\end{equation}
where $D$ is the diffusion constant. Indeed the error function appears any 
time there is a summation of the effects of a series of line sources each of 
which has an exponential distribution, both in finite and infinite media, as 
discussed in great detail in \cite{crank75}.


Further, the error function can be related to special values of the 
degenerate hypergeometric function, ${}_1F_1(\alpha;\gamma;z)$. In 
particular:
\[ {}_1F_1(\frac{1}{2};\frac{3}{2};-x^2) \simeq  \frac{\sqrt{\pi}}{2x} 
\tanh(\frac{2}{\sqrt{\pi}}x) \]

Our final example comes from the theory of parabolic cylinder functons, 
$D_p(z)$, which are solutions to the differential equation:
\begin{equation}
\frac{d^2 u}{dz^2} + (p + \frac{1}{2} - \frac{z^2}{4})u = 0
\end{equation}
with $u = D_p(z)$ and for integer values of $p = n$, they are related to 
the Hermite polynomials, $H_n(z)$ by $D_n(z) = 2^{-n/2} e^{-z^2/4} 
H_n(\frac{z}{\sqrt{2}})$. Finally we may write, for the special cases of 
$n = -1,-2$: 
\begin{eqnarray}
D_{-1}(z) &\simeq& e^{z^2/4} \sqrt{\frac{\pi}{2}} \left[1 - 
\tanh(\sqrt{\frac{2}{\pi}} z) \right]\\
D_{-2}(z) &\simeq& -e^{z^2/4} \sqrt{\frac{\pi}{2}} 
\left[\sqrt{\frac{2}{\pi}} 
e^{-z^2/2} - z (1 - \tanh(\sqrt{\frac{2}{\pi}}z)) \right]\\
\end{eqnarray}

\section{Conclusions}

In this {\em Letter} we have presented a function approximating 
$\mbox{erf}(x)$ to better than $4 \%~~\forall~~x$, with exponential 
convergence as $x \rightarrow \infty$. This solution is simply the kink 
soliton, $\phi(x) = \tanh(2x/\sqrt{\pi})$ and can be optimised for accuracy 
if the error function at small values of the argument is required.  

Further we have found a solution with maximum error of  $0.15\%$ by adding 
a generalised Maxwell distribution to the kink soliton, equations 
(\ref{eq:differr}), (\ref{eq:tot}).  Future 
work should be aimed at finding truly optimal solutions. Finally 
a few applications were discussed, particularly to diffusion dynamics and to 
the  generation of Gaussian random fields.  

The author would like to thank Prof. Domb, Claudio Scrucca and Lando Caiani 
for illuminating discussions and Stefano Bianchini for a very useful critical 
reading of the manuscript.

\end{document}